\documentclass[12pt,preprint]{aastex}

\usepackage{natbib}

\shorttitle{Steady X-Ray Synchrotron Emission in SN~1006}
\shortauthors{S. Katsuda et al.}

\begin{document}

\title{Steady X-Ray Synchrotron Emission in the Northeastern Limb 
  of SN~1006}

\author{Satoru Katsuda\altaffilmark{1}, Robert
  Petre\altaffilmark{1}, Koji Mori\altaffilmark{2}, 
 Stephen P. Reynolds\altaffilmark{3}, Knox S.
  Long\altaffilmark{4}, P. Frank Winkler\altaffilmark{5}, and Hiroshi 
  Tsunemi\altaffilmark{6}
}

\email{Satoru.Katsuda@nasa.gov, Robert.Petre-1@nasa.gov,
  mori@astro.miyazaki-u.ac.jp, reynolds@ncsu.edu,
  long@stsci.edu, winkler@middlebury.edu,
  tsunemi@ess.sci.osaka-u.ac.jp}

\altaffiltext{1}{NASA Goddard Space Flight Center, Greenbelt, MD
   20771, U.S.A.} 

\altaffiltext{2}{Department of Applied Physics, Faculty of Engineering,
University of Miyazaki, 1-1 Gakuen Kibana-dai Nishi, Miyazaki, 889-2192,
Japan}

\altaffiltext{3}{Physics Department, North Carolina State University,
  Raleigh, North Carolina 27695}  

\altaffiltext{4}{Space Telescope Science Institute, 3700 San Martin
  Dr., Baltimore, MD 21218, U.S.A.} 

\altaffiltext{5}{Department of Physics, Middlebury College,
  Middlebury, VT 05753}

\altaffiltext{6}{Department of Earth and Space Science, Graduate School
of Science, Osaka University,\\ 1-1 Machikaneyama, Toyonaka, Osaka,
560-0043, Japan}


\begin{abstract}

We investigate time variations and detailed spatial structures of
X-ray synchrotron emission in the northeastern limb of SN~1006, using
two {\it Chandra} observations taken in 2000 and 2008.  We extract
spectra from a number of small ($\sim$10$^{\prime\prime}$) regions.
After taking account of proper motion and isolating the synchrotron
from the thermal emission, we study time variations in the synchrotron
emission in the small regions.  We find that there are no 
regions showing strong flux variations.  Our analysis shows an
apparent flux decline in the overall synchrotron flux of $\sim$4\% at
high energies, but we suspect that this is mostly a calibration
effect, and that flux is actually constant to $\sim$1\%.  This is much
less than the variation found in other remnants where it was used to
infer magnetic-field strengths up to 1 mG.  We attribute the lack of 
variability to the smoothness of the synchrotron morphology, in 
contrast to the small-scale knots found to be variable in other
remnants.  The smoothness is to be expected for a Type Ia remnant
encountering uniform material.  Finally we find a spatial correlation 
between the flux and the cut-off frequency in synchrotron emission.
The simplest interpretation is that the cut-off frequency depends on
the magnetic-field strength.  This would require that the maximum
energy of accelerated electrons is not limited by synchrotron losses,
but by some other effect.  Alternatively, the rate of particle
injection and acceleration may vary due to some effect not yet
accounted for, such as a dependence on shock obliquity.

\end{abstract}

\section{Introduction}

SN~1006 is the supernova remnant (SNR) for which X-ray synchrotron
emission from diffusive shock accelerated electrons was first proposed 
(Reynolds \& Chevalier 1981) and detected in X-rays with {\it ASCA}
(Koyama et al.\ 1995).  It remains an unrivaled laboratory for
studying these phenomena because of its large size ($\sim30^\prime$ 
diameter; Winkler \& Long 1997) and low interstellar absorption 
(6.8$\times$10$^{20}$\,cm$^{-2}$; Dubner et al.\ 2002).  Very
recently, the HESS team reported the firm detection of TeV
$\gamma$-ray emission from this SNR (Acero et al.\ 2010).

One of {\it Chandra}'s great discoveries in particle acceleration
physics was that rims of SN~1006 and other young SNRs are very narrow,
much narrower than the 1/12 shock radii expected for a strong shock
with a compression ratio of 4 (e.g., Long et al.\ 2003; Bamba et al.\
2003; 2005).  There is a general consensus that these narrow filaments
are indirect evidence for strongly amplified magnetic fields at or
upstream of the shock.  However, the origin of the narrowness has been
debated; two interpretations proposed so far have clearly different
scenarios (e.g., Cassam-Chena\"i et al.\ 2007).  One considers the
effect of a rapid decay of the amplified magnetic field downstream, so
that bright narrow magnetic filaments are formed behind the shock
(Pohl et al.\ 2005).  The other assumes a relatively constant, strong
magnetic field downstream of the shock.  In this case, accelerated
electrons quickly lose their energy through synchrotron radiation,
resulting in narrow synchrotron X-ray filaments (e.g., V\"olk et al.\
2005 and references therein).  The nature of the magnetic-field
amplification is not well understood at this time; the post-shock
evolution of the field holds important clues to the process, so
settling the question of the mechanism limiting filament widths has
considerable significance.

Meanwhile, in RX~J1713.7-3946 and Cas~A SNRs, several
synchrotron-dominated knotty features, whose size is about
10$^{\prime\prime}$, were found to show year-scale time variations
(e.g., Uchiyama et al.\ 2007; Patnaude \& Fesen 2009).  The rapid
variations may reflect fast acceleration or cooling of accelerated
electrons in strongly amplified magnetic fields up to the level of mG.
On the other hand, more diffuse regions in 
these SNRs do not show rapid variations, which leads the authors to
consider that these regions have somewhat weaker magnetic fields.
Thus, it is now possible to roughly estimate magnetic-field strengths,
$B$, in SNRs, when time variations in synchrotron emission can be
measured.  Synchrotron X-ray flux variation can be produced
stochastically, even in the absence of variations in the electron
distribution, in the presence of a stochastic magnetic field (Bykov,
Uvarov, \& Ellison 2008). Somewhat smaller rms values
of magnetic field are required, but substantial amplification is still
necessary.  (It should be noted that absence of variation does
not demand low magnetic-field strengths; systematic, smooth
steady-state magnetic-field amplification could produce steady
emission varying only on overall SNR dynamical timescales, decades to
centuries.)

In this paper, we investigate time variations of discrete features
in the northeastern (NE) limb of SN~1006, using two {\it Chandra}
observations taken in 2000 and 2008.  We have recently measured proper
motions of the shock fronts in the NE limb to be almost uniform at 
0.$^{\prime\prime}$5\,yr$^{-1}$ (Katsuda et al.\  2009; hereafter 
Paper I).  By correcting for the proper motion, we can track the same 
regions away from the shock front (or loosely, the same fluid
elements) in two epochs.  We also reveal detailed spatial
structures of the synchrotron emission in the NE limb.  Based on the
results, we discuss why the synchrotron filaments are so narrow and
the mechanism that limits the maximum energy of accelerated electrons. 

\section{Observations}

We use two {\it Chandra} observations taken in 2000 (ObsID.\ 732) and
2008 (ObsID.\ 9107) that are the same data presented in our previous 
proper-motion measurements (Paper I).  We use the reduced data
products described in Paper I.  We note that the
second observation was specifically intended to allow a proper-motion 
measurement, and therefore the pointing direction, roll angle, and 
exposure time are the same as those in the first observation.  This
configuration was chosen to allow as precise a comparison in the two
epochs as possible because the same physical regions are seen at
almost the same detector position with the same effective area and
spatial resolution. 

\section{Analysis and Results}

\begin{figure}
\plottwo{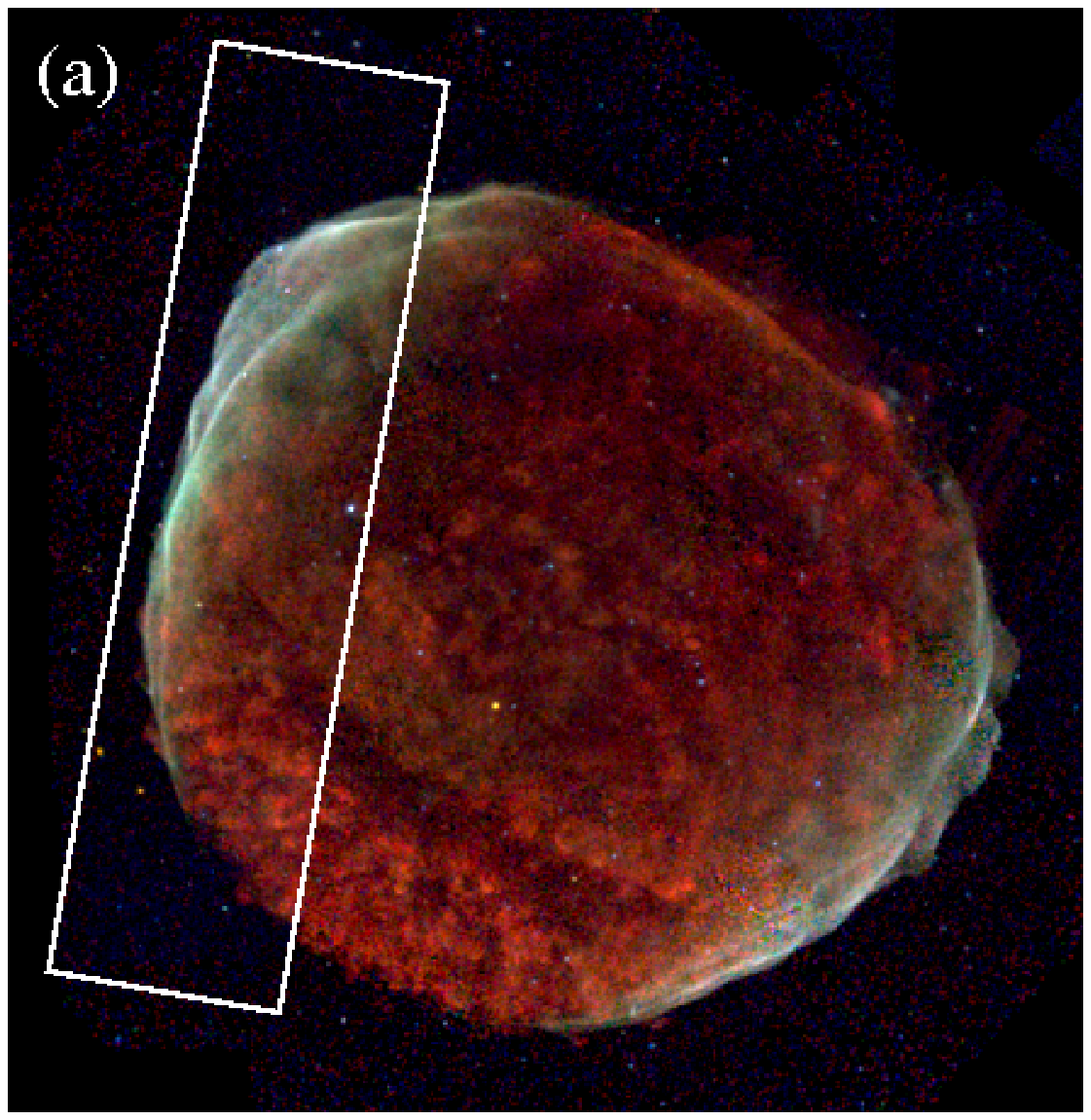}{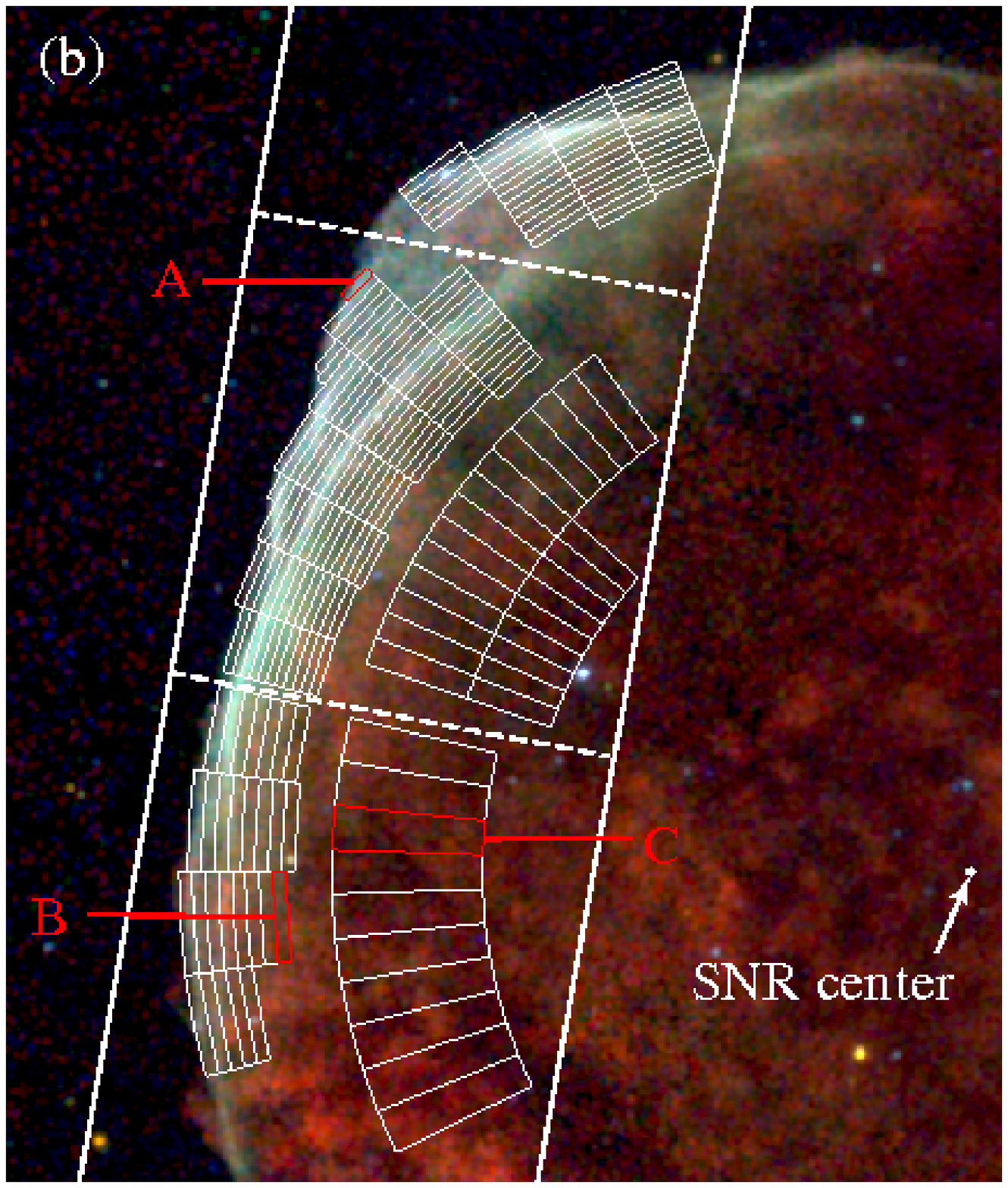}
\caption{(a) {\it Chandra} three-color image after vignetting
  effects are corrected.  Red, green, and blue correspond to
  0.5--0.8\,keV (mostly, K lines of O), 0.8--2.0\,keV (mostly, K lines 
  of Ne, Mg, and Si), and 2.0--5.0\,keV (mostly, synchrotron emission) 
  bands, respectively.  The image is 
  binned by 1$^{\prime\prime}$.97 and has been smoothed by a Gaussian
  kernel of $\sigma = 5^{\prime\prime}.90$.  The intensity scale is
  square root.  The field of view of the {\it Chandra} observations
  of the NE limb (ObsIDs 732 and 9107) are shown as a white box.  
  (b) Same as Fig.~\ref{fig:image} (a), but focused on the NE limb. 
  The field of view the {\it Chandra} observations are within two
  white lines.  We extract spectra from white (and red) pie-shaped
  regions.  Example spectra for red regions indicated as letters A--C
  are shown in Figs.~\ref{fig:spec1} and \ref{fig:spec2}.  The SNR
  center of [(ra, dec) =
  $15^\mathrm{h}$02$^\mathrm{m}$54$^\mathrm{s}$.9, 
  $-41^{\circ}56^{\prime}08^{\prime\prime}.9$ (J2000)] is taken from 
  Paper I. 
} 
\label{fig:image}
\end{figure}

Figure~\ref{fig:image} (a) shows a three-color {\it Chandra} image of 
SN~1006, where red, green, and blue correspond to 0.5--0.8\,keV 
(mostly, K-shell lines of O), 0.8--2.0\,keV (mostly, Ne, Mg, and Si K
lines), and 2.0--5.0\,keV (mostly, synchrotron continuum) bands,
respectively.  Regions seen in white are dominated by nonthermal
synchrotron emission, while those in red or green are dominated by
thermal emission.  In this paper, we focus on the nonthermal emission,
for which we will investigate time variations as well as detailed 
spatial structures. 

As shown in Fig.~\ref{fig:image} (b), we extract spectra from a number
of small regions that are annular sectors covering the
nonthermally-dominated area in the NE limb.  We use the SNR center of
[(ra, dec) = ($15^\mathrm{h}$02$^\mathrm{m}$54$^\mathrm{s}$.9,
$-41^{\circ}56^{\prime}08^{\prime\prime}.9$) (J2000)] determined from
the {\it ROSAT} HRI image (Paper I).  The sizes of the regions range
from 10$^{\prime\prime}\times30^{\prime\prime}$ to
15$^{\prime\prime}\times115^{\prime\prime}$ to assure that 
each regions contains about 3000 counts. There are 175 regions in
all.  For simplicity of our spectral analysis, we exclude boundary
regions between the front- and back-illuminated chips (the boundaries
are indicated as dashed lines in Fig.~\ref{fig:image} (b)).  Since we
know that forward shocks in the NE limb are moving at
$0^{\prime\prime}.5$\,yr$^{-1}$ (Paper I), we simply shift all the
regions by 4$^{\prime\prime}$ outward for the second observation.  In
this way, we extract two spectra (taken in 2000 and 2008) from each
region.  We subtract background emission from the source-free areas in
the identical chip of the same observation.  (The background never
amounts to more than 15\% of the total counts in a region, and is
normally much less, so the use of $\chi^2$ statistics is a reasonable
approximation.)  In addition to the X-ray data, we also use a VLA
image at 1.37\,GHz (Dyer et al.\ 2009) to constrain the normalization
of nonthermal emission.  We calculate radio fluxes from co-spatial
regions, i.e., the small regions shifted by 2$^{\prime\prime}$ outward
compared with those for the first observations, since the radio image
was taken in 2004.  The radio resolution is $14^{\prime\prime} \times
6^{\prime\prime}$ (long axis N-S), so the effect of this correction
should be small.

Although most of the regions show featureless spectra, some regions
exhibit K lines from metals such as O, Ne, Mg, or Si.  We thus employ
an absorbed nonthermal plus thermal components model, where we employ
the {\tt tbabs} model (Wilms et al.\ 2000) for absorption, the {\tt
srcut} model, which describes synchrotron emission from a power-law
distribution of electrons with an exponential cut-off (Reynolds \&
Keohane 1999), with correction described in Reynolds 2008 for the
nonthermal component, and the {\tt vpshock} model, which describes
thermal emission from a non-equilibrium ionization (NEI) plane-shock
plasma, in conjunction with NEI version 2.0 (Borkowski et al.\ 2001)
for the thermal component.  In the fitting, photons in an energy range
of 0.4--8.0\,keV are used.  We fix the intervening hydrogen column
density, $N_\mathrm{H}$, to be 6.8$\times$10$^{20}$cm$^{-2}$ (Dubner
et al.\ 2002).  In the {\tt vpshock} component, we fix the abundances
of O, Ne, Mg, and Si to be 4.4, 1.5, 15, and 50 times solar values
(Anders \& Grevesse 1989), respectively, following the most recent
{\it XMM-Newton} results (Miceli et al.\ 2009).  Other elemental
abundances are fixed to solar values.  The electron temperature,
$kT_\mathrm{e}$, and the ionization timescale, $n_\mathrm{e}t$, are
fixed to 0.5\,keV and $1\times10^{10}$\,cm$^{-3}$\,sec, respectively
(Miceli et al.\ 2009), where $n_\mathrm{e}t$ is the electron density
times the elapsed time after shock heating and the {\tt vpshock} model
assumes a range of $n_\mathrm{e}t$ from zero up to
$1\times10^{10}$\,cm$^{-3}$\,sec. Note that Miceli et al.'s results
actually show that $kT_\mathrm{e} \sim 0.4$\,keV and $n_\mathrm{e}t
\sim 1.5 \times 10^{10}$\,cm$^{-3}$\,sec in the NE limb, but this
parameter set does not affect our spectral-fit parameters presented
below.  The only free parameter we set in the {\tt vpshock} model is
the volume emission measure (VEM; VEM $=\int n_\mathrm{e}n_\mathrm{H}
dV$, where $n_\mathrm{H}$ is the number density of protons, and $V$ is
the X-ray--emitting volume).  For the {\tt srcut} component, we let
the cut-off frequency and the mean spectral index (the $\alpha$
parameter) inferred from the X-ray spectrum be free parameters, 
whereas the normalization (the flux at 1\,GHz) is fixed to the value
extrapolated from the radio flux at 1.37\,GHz (Dyer et al.\ 2009)
assuming a photon index of 0.55.  In the 
initial fits, we allowed cut-off frequencies to vary freely in the two
(2000 and 2008) data sets, but we found them to be consistent with
each other.  We thus simultaneously fit the 2000 and 2008 spectra, by
linking all the spectral-fit parameters except for an additional
parameter, the relative intensity of the {\tt srcut} component between
2000 and 2008, which is allowed to vary freely so that we can measure
time variations in its flux.

\begin{figure}
\plottwo{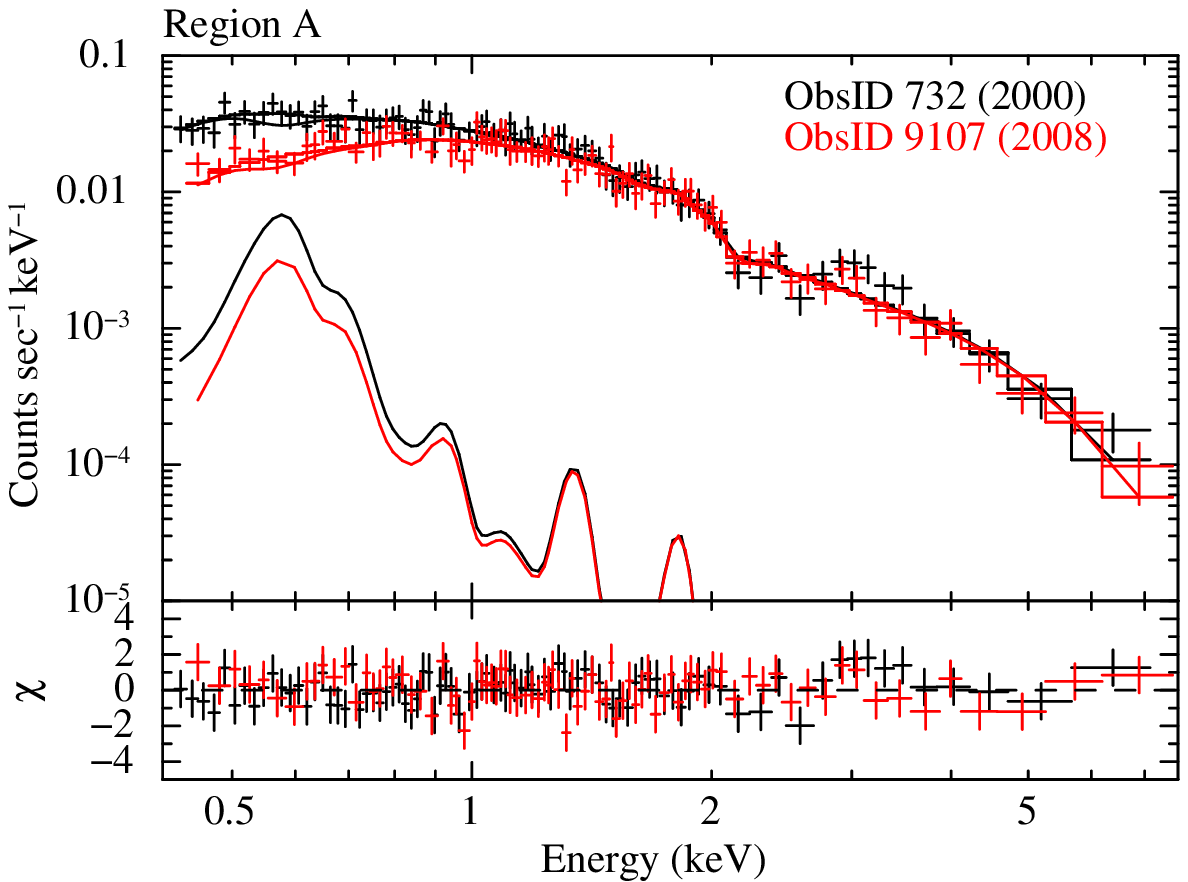}{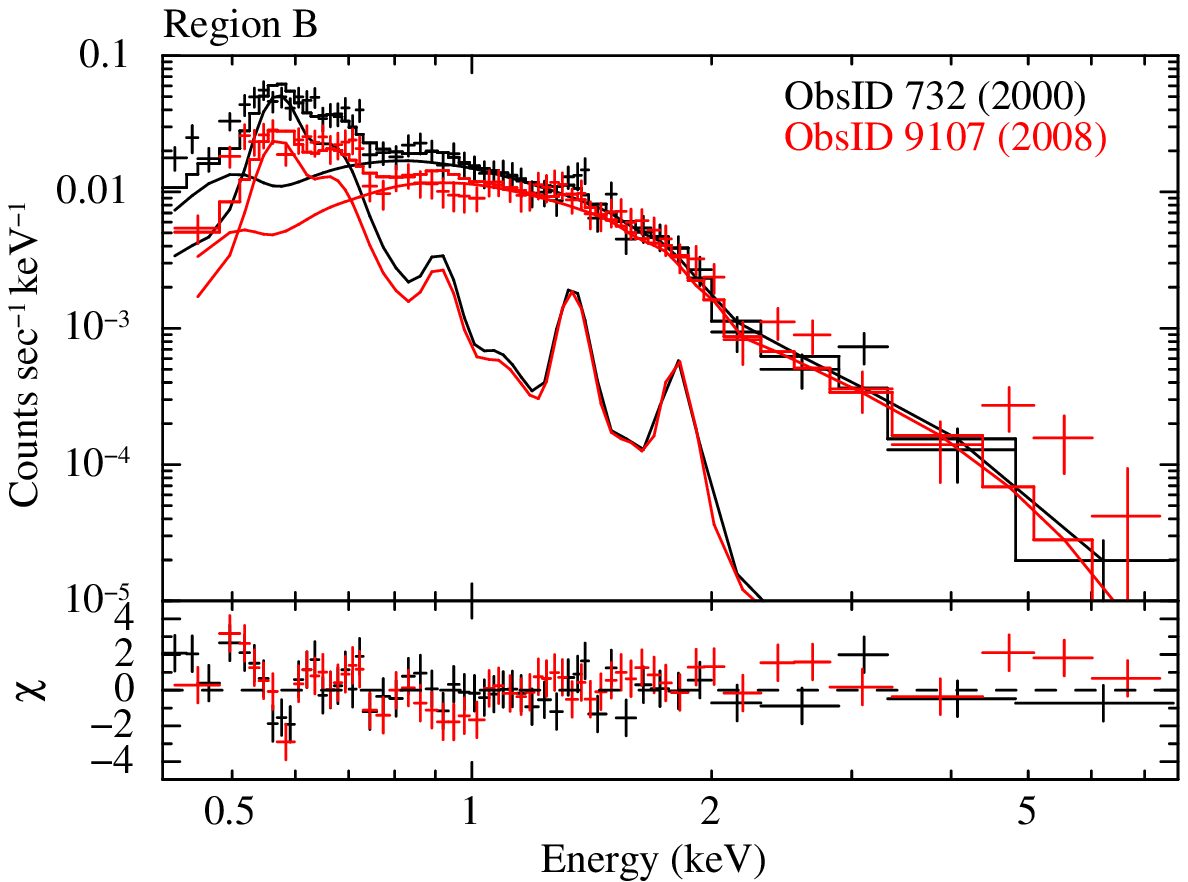}
\caption{Left: Example spectra extracted from region A indicated in
  Fig.~\ref{fig:image} along with the best-fit models and the
  residuals.  Black and red correspond to the 2000 and 2008 data,
  respectively.  Components of nonthermal and thermal emission are
  separately illustrated.  Right: Same as left but for region B
  indicated in Fig.~\ref{fig:image}. 
} 
\label{fig:spec1}
\end{figure}

Figure~\ref{fig:spec1} shows example spectra extracted from regions A
(with no significant thermal emission) and B (with significant thermal
emission) indicated in Fig.~\ref{fig:image} (b).  Black and red
correspond to 2000 and 2008, respectively.  The spectral difference
between the two colors, which is clearly seen below 1\,keV, shows the
accumulation of molecular contaminants on the ACIS-S optical blocking
filter.  Also shown in the figure are the best-fit models and the
residuals. Since the evolution of the contaminants is accounted for in
the response files, the same model (with slightly adjusted intensity
of the {\tt srcut} component) fits both 2000 and 2008 data well.
Spectral-fit parameters and fit statistics for the example spectra are
summarized in Table~\ref{tab:param}.  In this fitting procedure, a
number of parameters for the thermal component are assumed and fixed.
Although these values are plausible, it is worth checking the
sensitivity of the fit results to varying these parameters.  Before
investigating, we first note that fit results for most of the regions
are not sensitive to the treatments of thermal parameters since these
regions are dominated by nonthermal emission like Region A
(Fig.~\ref{fig:spec1} left).  On the other hand, the rest of the
regions including Region B (Fig.~\ref{fig:spec1} right), where
contributions of the thermal emission are relatively large, could be
affected by the assumptions in the fitting.  We tried fitting the
spectrum from Region B with $\pm$10\% different values for
$N_\mathrm{H}$, $kT_\mathrm{e}$, and $n_\mathrm{e}t$.  We found that
the fit results, i.e., the best-fit parameters in the {\tt srcut}
component, are not significantly changed from the original results
listed in Table~\ref{tab:param}.  Thus, variations of up to 10\% in
these parameters would not affect our results.  We have also checked
different sets of metal abundances.  We examined three cases: (1)
C=N=O=4.4, (2) Si=S=50, and (3) Fe=(Ni=)20 times the solar values,
where the Fe abundance is based on the Fe-rich ejecta measured in the
southeastern portion of SN~1006 with {\it Suzaku} (Yamaguchi et al.\
2008).  The first case yields a slightly better fit than the original
fit, but the best-fit parameters are consistent with those in
Table~\ref{tab:param}.  The second case gave us almost the same
results as those in Table~\ref{tab:param}, since S K lines are
negligible for the assumed plasma conditions of ($kT_\mathrm{e}$,
$n_\mathrm{e}t$)=(0.5\,keV, 1$\times10^{10}$\,cm$^{-3}$\,sec).  In the
third case, we find significantly different results: the photon index
and the cut-off frequency are found to be 0.534$\pm0.007$ and
2.3($\pm0.2)\times10^{16}$\,Hz, respectively.  However, the fit level
($\chi^2$=158) is not as good as that in Table~\ref{tab:param}.
Therefore, we believe that the Fe abundance in the NE limb is more
likely to be closer to the solar values rather than 20 times the solar 
values, and that our fitting procedure with the solar abundance for Fe 
is robust.

\begin{figure}
\includegraphics[angle=0,scale=0.65]{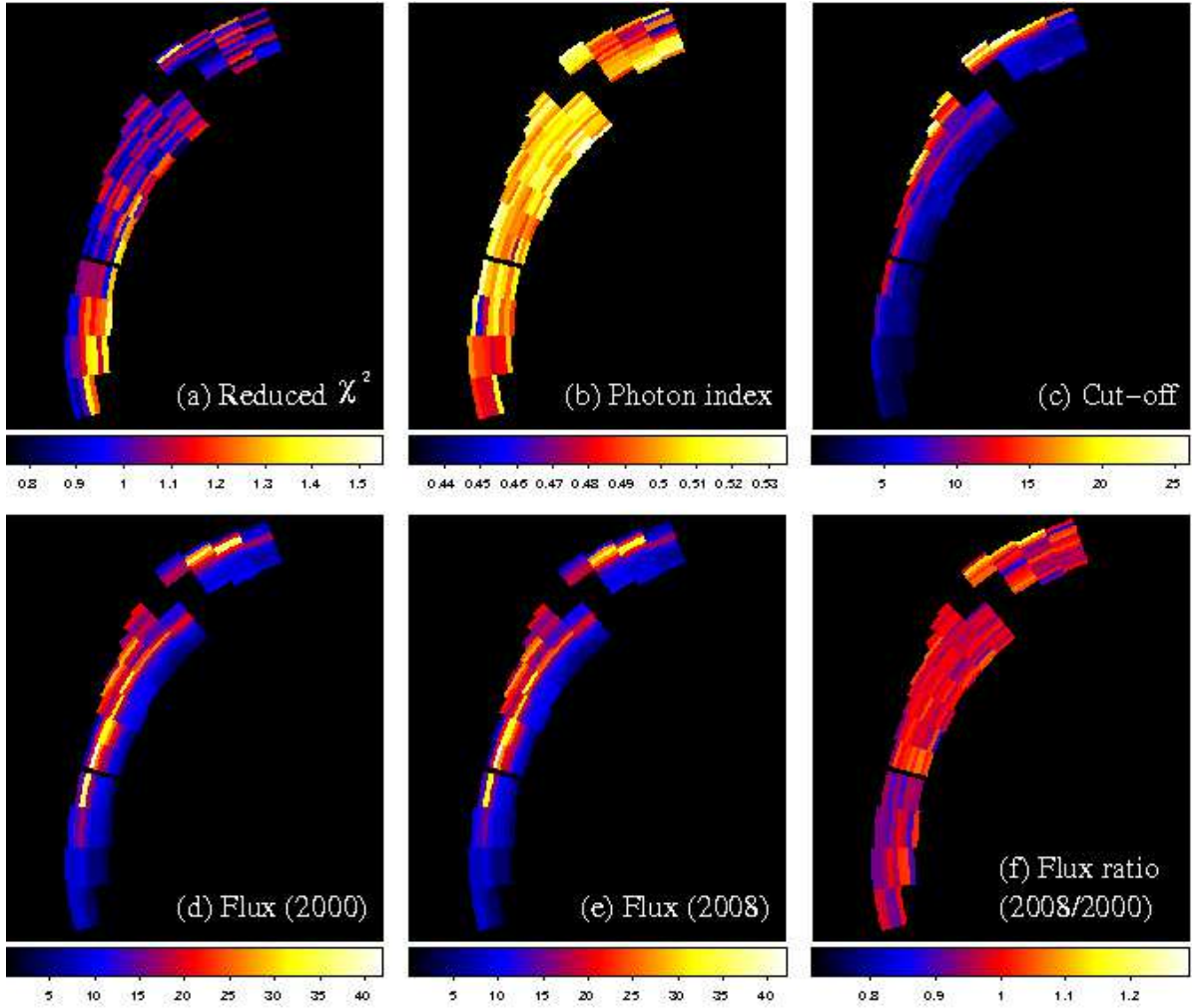}\hspace{1cm}
\caption{Results from the spatially-resolved spectral analysis for the 
  nonthermally-dominated regions.  Panels (a)--(f) show distributions of
  reduced $\chi^2$s, mean spectral indices inferred from the X-ray spectra, 
  cut-off frequencies, fluxes corrected for the interstellar absorption
  in 0.4--8.0\,keV for 2000 and 2008, and the flux ratios,
  respectively.  Values of the cut-off frequencies and fluxes are in
  units of 1$\times$10$^{16}$\,Hz and
  1$\times$10$^{-13}$\,ergs\,cm$^{-2}$\,sec$^{-1}$\,arcmin$^{-2}$,
  respectively.
} 
\label{fig:spatial}
\end{figure}

Maps of the the reduced $\chi^{2}$s, best-fit parameters (mean
spectral index and cut-off frequencies), fluxes in the {\tt srcut}
component (2000 and 2008), and the flux ratios (2008/2000) are shown
in Fig.~\ref{fig:spatial} (a)--(f), where fluxes are calculated in the 
0.4--8.0\,keV band after correcting for interstellar absorption.
Figure~\ref{fig:spatial} (a) shows that the fits are fairly good for
all the regions: the reduced $\chi^2$s are derived to be less than
1.5.  We see relatively worse fits at the southern regions.  This
is because of the simplicity of the thermal model.  Relatively large
residuals are found around 0.7\,keV energies where spectral modeling
of the thermal emission is quite hard due to either missing K lines of
O or inadequate atomic data for Fe L-shell lines (e.g., Yamaguchi et
al.\ 2008).  Also, such residuals are particularly evident in the 
southern regions where the contributions of thermal emission are
relatively large compared with the rest of the regions.  It is highly
likely that this discrepancy does not affect the spectral-fit
parameters in the nonthermal component.  Therefore, we are confident
of the best-fit parameters shown in Fig.~\ref{fig:spatial}.

As shown in Fig.~\ref{fig:spatial} (b), mean spectral index are
inferred to be around 0.5, which is consistent with the recent results 
from {\it Chandra} (Allen et al.\ 2008) and {\it XMM-Newton} (Miceli 
et al.\ 2009), but is slightly flatter than the radio value of
0.60 (0.51--0.68 for 90\% C.L.) reported for the integrated spectrum
(Allen et al.\ 2008).  Our inferred mean spectral indices depend on
flux ratios between radio and X-rays.  Thus, systematic uncertainties 
of radio fluxes, which could originate from relatively worse spatial
resolution of the radio image than that of the X-ray image, are
subject to additional uncertainties of the mean spectral index.  We 
check for the Region~A spectrum that a 50\% larger radio flux would
yield a $\sim$5\% larger photon index.  Therefore, together with the
relatively large uncertainty ($\sim$15\%) of the radio value from the
integrated  spectrum (Allen et al.\ 2008), we do not formally find
significant inconsistency of the mean spectral index.  Nonetheless,
the best-estimated values show discrepancy, which would suggest a
curved nonthermal spectrum expected in a nonlinear theory of diffusive
shock acceleration (Reynolds \& Ellison 1992), as previously noted by
others (e.g., Allen et al.\ 2008).  Cut-off frequencies in
Fig.~\ref{fig:spatial} (c) show strong  variations in both radial and 
azimuthal direction.  These spatial variations are also generally
consistent with previous studies (Rothenflug et al.\ 2004; Allen et
al.\ 2008; Miceli et al.\ 2009). We find a correlation between the
cut-off frequency and the flux, for the first time.  This will be
briefly discussed in the next section. Detailed discussion about these
spatial structures will be published elsewhere.

\begin{figure}
\includegraphics[angle=0,scale=0.65]{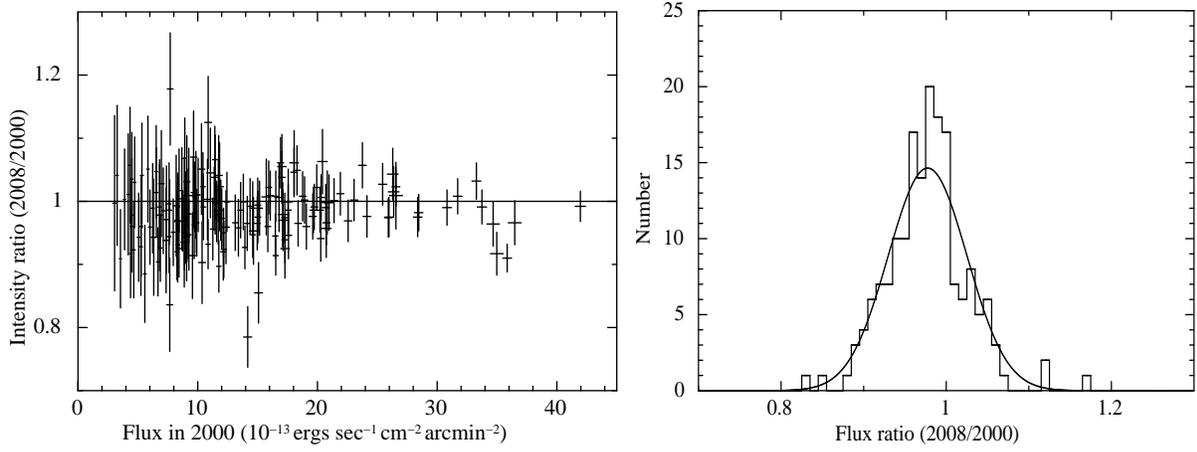}\hspace{1cm}
\caption{Left: Flux ratio in the srcut component as a function of the
  flux in 2000.  Errors of the vertical axis are 90\% confidence
  ranges of the constant parameters for the {\tt srcut} component,
  while those of the horizontal axis are square root of photon
  numbers in the srcut component.  A horizontal solid line drawn in the
  figure represent a constant flux line. 
  Right: Histogram of flux ratios (2008/2000).  The best-fit Gaussian 
  function is shown together.
} 
\label{fig:hist1}
\end{figure}

We find flux maps of 2000 and 2008 to be quite similar to each 
other.  This is confirmed by a flux ratio map in
Fig.~\ref{fig:spatial} (f), where most of the regions are in red
(i.e., constant fluxes).  In the figure, we see that a few northern
regions show somewhat higher values (yellow color in 
Fig.~\ref{fig:spatial} (f)) than the others.  This is because 
of imperfect correction for the proper motions there; these regions 
have slightly larger proper motions than assumed here (see Paper 
I) and we checked that, if we choose spectral extraction regions more 
explicitly, we do not see flux variations.  To study the flux 
variation more quantitatively, we plot the flux ratio (2008/2000) 
as a function of the flux in 2000 in Fig.~\ref{fig:hist1} left.  We 
can see that the data points are clustered tightly about a line 
representing no time variation.  This means that the fluxes in 2000
and 2008 are quite similar with each other for most of the regions.
On the other hand, it should be noted that there are a few data points
showing relatively large variations of $\sim$20\%.  We can not rule
out dramatic changes in those otherwise completely undistinguished
regions, but we believe that the fluxes in these regions are not really
changing.  This is because some of them are expected to be due to 
imperfect corrections for the proper motions as mentioned above.  And
others are just statistical fluctuations, because (1) we have checked
that these regions are randomly scattered (Fig.~\ref{fig:spatial} (f))
and (2) the histogram of the flux ratio is well represented by a
Gaussian function as shown in Fig.~\ref{fig:hist1} right.  In fact,
if we apply a constant model to the flux ratios in
Fig.~\ref{fig:hist1} left, we obtain a fairly good fits of
$\chi^2$/dof = 157/174 with a null hypothesis probability of 0.82.   
Thus, from a statistical point of view, we cannot reject the
possibility that the fluxes are constant everywhere in the NE limb.
Moreover, by looking at spectra from these regions, we see
nothing special in their spectral features nor particular spectral 
parameters.  In this context, we conclude that there are no peculiar
regions showing strong time variations, and that all the regions show
little or no time variations.  We notice that the center of the
histogram in Fig.~\ref{fig:hist1} right is apparently shifted from
unity: the Gaussian center is measured to be 0.977$\pm$0.006 (90\%
C.L.).


The apparent decline in the global synchrotron flux is interesting,
but falls within the calibration uncertainties for the effective area
(3\%\footnote{http://web.mit.edu/iachec/IACHEC\_2\_talks/IACHEC\_II\_chandra\_summary.pdf})
determined by the CXC.  Therefore, we have explored a variety of
possibilities for improving the relative calibration of our 
measurement.  There are relatively few point sources in the field, and
these are not expected to be constant with time in any event. Indeed,
the brightest source, QSO1 in Winkler et al.\ (2005) was brighter in
2008 by 50\% than in 2000.  A more promising alternative is to search 
for a change in the thermal emission and compare this to changes in
the nonthermal emission.

Our strategy to estimate changes in nonthermal and thermal emission is
as follows.  We first estimate nonthermal flux variations from the
nonthermally-dominated regions (i.e., the outer regions elongated in
the azimuthal direction in Fig.~\ref{fig:image} (b)) as we have
already analyzed.  In this process, we assume no time variation for
thermal emission to avoid possible degeneracy in separating thermal
and nonthermal components; it is difficult to estimate the
contribution of thermal emission correctly in these regions, and
incorrect intensity ratios between the two epochs for thermal emission
would affect those for nonthermal emission more or less.  In any case,
thermal emission makes only a small contribution in those regions.
Next, we estimate thermal flux variations from thermally-dominated
regions (i.e., interior regions elongated in the radial direction in
Fig.~\ref{fig:image} (b)).  (Here, we assume the flux variation of
nonthermal emission has the value measured above.)  Spectra of
these thermally-dominated regions are clearly distinct from those of
the nonthermally-dominated regions (Fig.~\ref{fig:spec1}),
resulting in different mean photon energies between them.  To compare
fluxes in these different kinds of spectra as accurately as possible,
it may be important to consider the energy dependence of effective area
and quantum efficiency.  Therefore, we divide the spectra into three
energy bands: 0.4--0.8\,keV (K lines of O), 0.8--1.0\,keV (K lines of
Ne), and 1.0--8.0\,keV.  Finally, we compare flux variations between
nonthermal and thermal emission in the three energy bands.

\begin{figure}
\plottwo{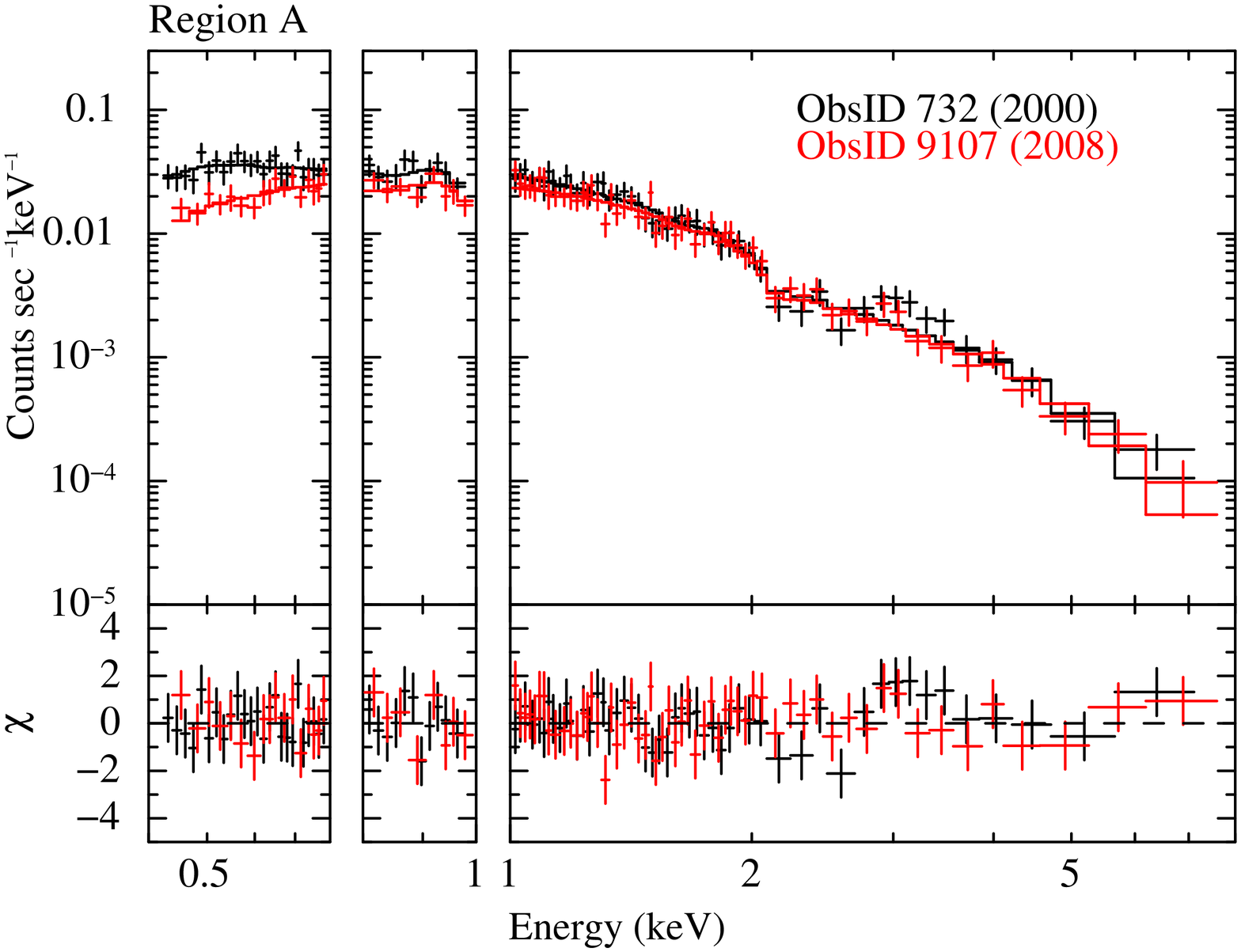}{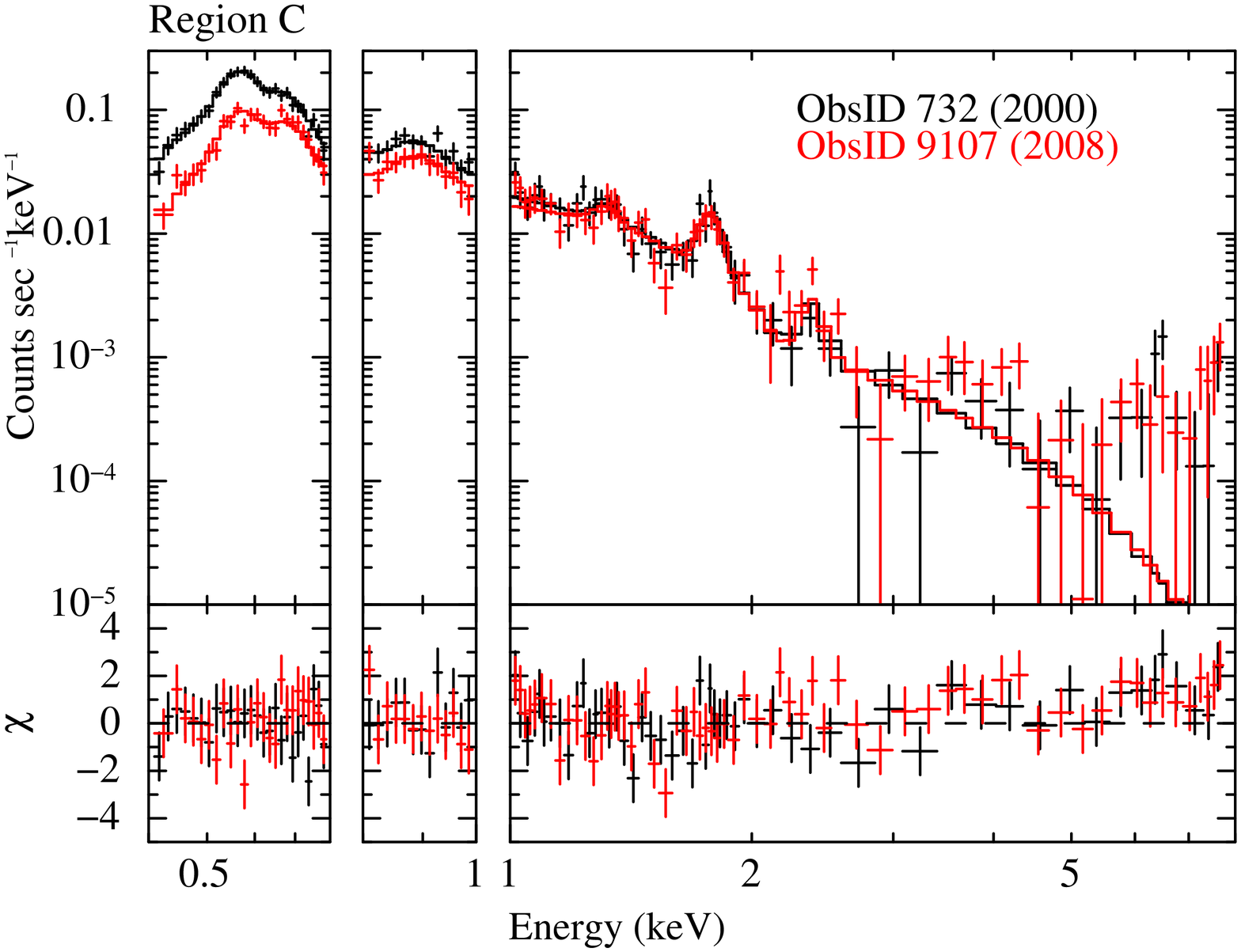}
\caption{Left: Same as Fig.~\ref{fig:spec1} left, but the spectra are
  fitted separately in energy bands of 0.4--0.8\,keV, 0.8--1.0\,keV,
  and 1.0--8.0\,keV.
  Right: Same as left but for a thermally-dominated region,  
  region C indicated in Fig.~\ref{fig:image}.
}
\label{fig:spec2}
\end{figure}

\begin{figure}
\includegraphics[angle=0,scale=0.5]{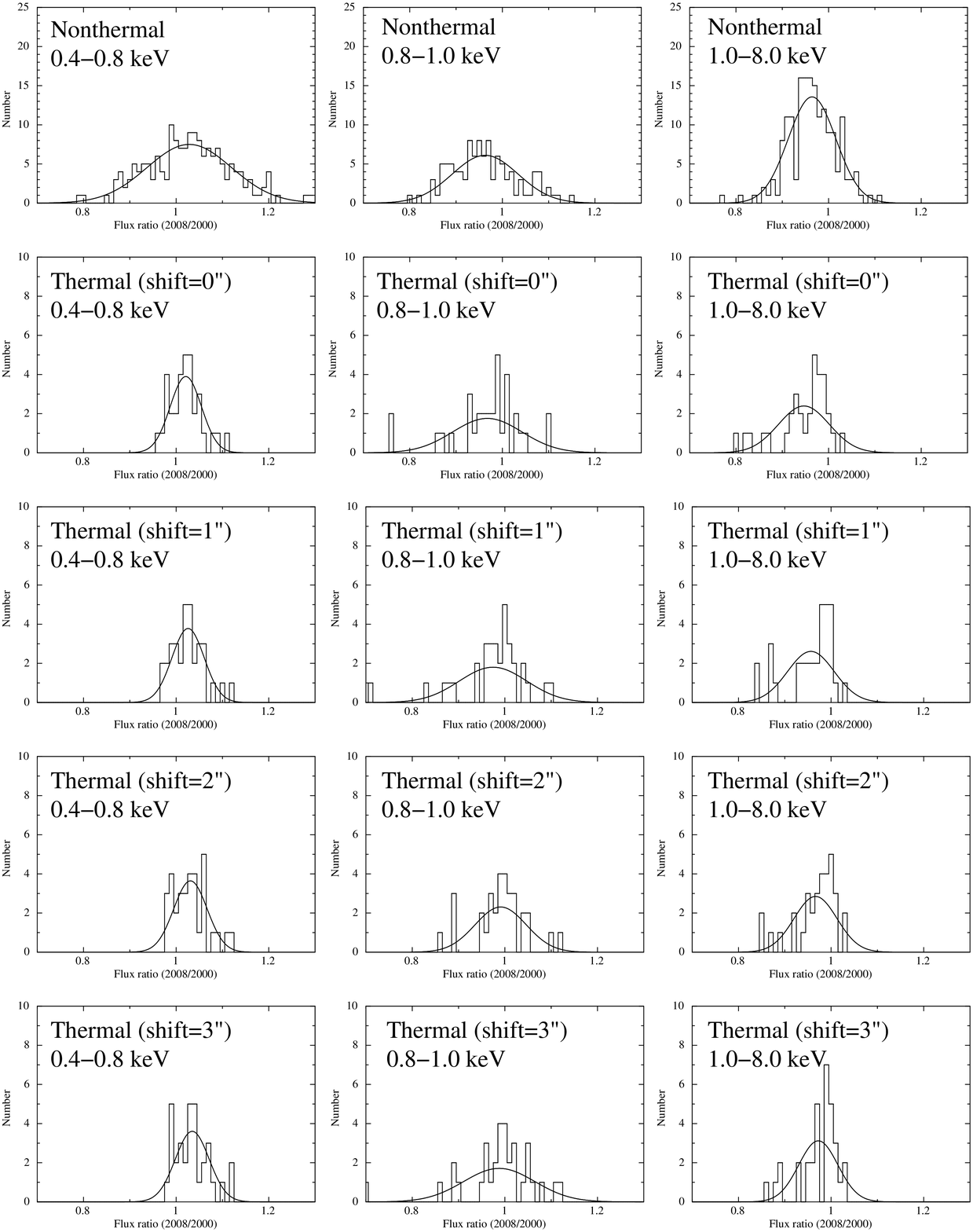}\hspace{1cm}
\caption{Histograms of flux ratios (2008/2000) for energy bands of
  0.4--0.8\,keV (left column), 0.8--1.0\,keV (middle column), and
  1.0--8.0\,keV (right column).  The first row is responsible for
  nonthermally-dominated regions, while others are responsible for 
  thermally-dominated regions. 
} 
\label{fig:hist_all}
\end{figure}

To measure flux variations in synchrotron emission of the three
energy bands separately, we re-fit spectra from nonthermally-dominated
regions.  We employ the same model used above (i.e., {\tt vpshock}
plus {\tt srcut}).  We also treat the spectral-fit parameters in the
same manner as above.  The only exception is that the mean spectral
indices are fixed to 0.5 which is typical in the NE limb (typical, if
no spectral curvature is assumed) for relatively narrow energy bands
of 0.4--0.8\,keV and 0.8--1.0\,keV where we cannot constrain both the
photon index and the cut-off frequency (Note that, when fitting the 
1.0--8.0\,keV band, we allow the mean spectral indices to vary freely,
since we can constrain them).   As mentioned above, we assume no time
variations for the thermal component.  In this way, we fit all the
spectra in the three energy bands, and derive statistically acceptable
fits for them. Example spectra from region A indicated in 
Fig.~\ref{fig:image} right are shown in Fig.~\ref{fig:spec2} left.
Then, fluxes of the nonthermal component are calculated from the
best-fit models.  Figure~\ref{fig:hist_all} (the first row) shows
histograms of the flux ratios (2008/2000) for the three energy bands
together with their best-fit Gaussian functions.  The best-fit values
of the Gaussian center are summarized in Table~\ref{tab:flux_ratio},
from which we can see the energy dependence of the flux variations.

Next, we investigate flux variations in thermal emission.  Since we
were concerned that the thermally-dominated regions might not have the 
same proper motion as the synchrotron dominated regions, we examine
four cases: 0$^{\prime\prime}$, 1$^{\prime\prime}$,
2$^{\prime\prime}$, and 3$^{\prime\prime}$ shifts in radial direction
between the two epochs. 
It is also difficult to satisfactorily reproduce the thermal emission
by plasma models (e.g., Yamaguchi et al.\ 2008).  Therefore, we 
alternatively apply a phenomenological model consisting of several
Gaussian components in addition to two bremsstrahlung components plus
an {\tt srcut} component.  The use of a phenomenological model is also 
justified by the fact that we are not trying to draw inferences from 
the model parameters but are just trying to get a good flux measurement.
For the 0.4--0.8\,keV band, we include five Gaussians at
$\sim$0.44\,keV (N He$\alpha$), $\sim$0.5\,keV (N Ly$\alpha$),
$\sim$0.57\,keV (O He$\alpha$), $\sim$0.66\,keV (O Ly$\alpha$ and O
He$\beta$), and $\sim$0.7\,keV (O He$\gamma$ and/or Fe L).  For the
0.8--1.0\,keV band, we include two Gaussians at $\sim$0.71\,keV (O
He$\delta$) and $\sim$0.91\,keV (Ne He$\alpha$).  For the
1.0--8.0\,keV band, we include three Gaussians at $\sim$1.35\,keV (Mg
He$\alpha$), $\sim$1.8\,keV (Si He$\alpha$), and $\sim$2.4\,keV (S
He$\alpha$).  Center energies, widths, and normalizations in the
Gaussian components are treated as free parameters, but for those at
0.7\,keV and 0.71\,keV, only normalizations are allowed to vary freely
with fixed center energies and fixed widths at zero.  We fix
$kT_\mathrm{e}$s in the two 
bremsstrahlung components to 0.5\,keV and 2.0\,keV, based on recent
X-ray analyses from {\it Suzaku} (Yamaguchi et al.\ 2008) and {\it
  XMM-Newton} (Miceli et al.\ 2009).  In the {\tt srcut} model, the
mean spectral index is fixed to 0.5. The normalization is also fixed to
the value estimated from the 1.37\,GHz image.  The cut-off frequency
is left as a free parameter.  The relative intensity of the {\tt
srcut} model between the two epochs is fixed to those derived in the
previous paragraph (see, Table~\ref{tab:flux_ratio}), whereas that of
the thermal component (i.e., the sum of all the components excluding
the {\tt srcut} component) is allowed to vary freely so that we can
obtain its flux variation.  This model yields statistically
acceptable fits for all the spectra in the three energy bands.
Example spectra from region C indicated in Fig~\ref{fig:image} right
are shown in Fig.~\ref{fig:spec2} right.  Similarly to the
nonthermally-dominated regions, we generate flux-ratio histograms of
the thermal component as shown in Fig.~\ref{fig:hist_all}, where the
second, third, fourth, and fifth rows are responsible for
0$^{\prime\prime}$, 1$^{\prime\prime}$, 2$^{\prime\prime}$, and
3$^{\prime\prime}$ shifted cases, respectively.  The values of the
best-fit Gaussian centers are summarized in
Table~\ref{tab:flux_ratio}.

Looking at Table~\ref{tab:flux_ratio}, we see that there are flux
changes in both the thermal and nonthermal emission and that the
changes in the two components.  In fact, ratios of the flux variations 
between nonthermal and thermal emission are calculated to be about
unity at all the three energy bands as also shown in
Table~\ref{tab:flux_ratio}.   This is strong evidence that the changes
in flux are due to calibration effects. 

Can the flux changes in both the thermal and nonthermal emission be
understood?   The fluxes increase by $\sim$3\% at low energies whereas 
they decrease by $\sim$4\% at high energies.   Also notable in the
table is energy dependence of flux variations: the fluxes increase by
$\sim$3\% at low energies, whereas they decrease by $\sim$4\% at high
energies.  Since the contaminants on the optical blocking filter could
influence spectra below 1\,keV, the increasing flux at low energies
could be due to this effect.  On the other hand, it cannot fully
explain the decreasing flux at high energies.

There are two possibilities for the variations at high energies.  One
is that some calibration effects cause the apparent changes for both
thermal and nonthermal emission, i.e., the fluxes are actually almost
constant with time.  In this case, the time variation of nonthermal
emission would be less than 1\% over 8\,yrs, based on the
time-variation ratio between nonthermal and thermal measured in
1--8\,keV (see, Table~\ref{tab:flux_ratio}).  This interpretation is
supported by the fact that flux variations of thermal emission are in
good agreement with those of nonthermal emission; it is likely that
the agreements are not just coincidence but that they have the same
underlying origin.  As an additional check of calibration effects, we
compared two observations of clusters of galaxies, since they are not
expected to change over the time period of interest (i.e.,
$\sim$10\,yrs).  We chose the Fornax cluster and HCG62, since they
were observed twice over this time period with the same chip (i.e.,
chip7) on the ACIS-S array.  We found that both of them are apparently
declining: $\sim$7\% between 2000 and 2009 for the Fornax cluster and
$\sim$3\% between 2000 and 2008 for HCG62.  This result implies the
presence of calibration effects.  Given that we measure relative
fluxes between 2000 and 2008, we need time-dependent calibration
effects to explain the flux variations seen.  These effects includes
the buildup of the ACIS contaminant, the increase in charge transfer 
inefficiency which could result in $\lesssim$1\% uncertainty in flux
measurements, and the variable particle background which could result
in $\sim$1\% uncertainty in flux measurements (a private communication
with Paul Plucinsky).  To estimate the flux uncertainty from
contaminants, we use the {\tt acisabs} model in XSPEC with response
files without corrections for the effects of contaminants.  This model
allows us to examine various amounts of contaminants by specifying
various time since the launch of {\it Chandra} in its parameter.  We
find that a 10\% variation of the quantum efficiency at 0.67\,keV
would result in a 1\% variation of the flux in 1--8\,keV for a
nonthermally-dominated spectrum.  Therefore, calibration uncertainties
in relative fluxes could be as large as $\sim$3\%, consistent with our 
measurements.  We conclude that the time variation in the flux from
synchrotron emission is most likely constant to 1\%, and certainly
less than 1\%.

However, we cannot fully rule out the other possibility that both
thermal and nonthermal emission are declining at similar rates by
chance.  Therefore, it is interesting to investigate the time
variation from a theoretical point of view.  Simple models for the
evolution of synchrotron brightness of SNRs (e.g., Reynolds \&
Chevalier 1981) predict the rate at which synchrotron flux  
should be dropping.  As shown in the Appendix A, assuming as in that
paper that both magnetic-field energy density and
relativistic-electron energy density scale with postshock pressure $P
\propto \rho u_s^2$, but generalizing from the assumption of Sedov
evolution made there to the observed expansion rate $R \propto t^m$
with $m = 0.54$ (Paper I), we predict that above the cut-off frequency
$\nu_{\rm cutoff}$ the synchrotron intensity should drop off at (0.2
-- 0.25)\% yr$^{-1}$ (1.6\% -- 2.0\% between 2000 and 2008), of the
same order as the small variation we find.  The thermal emission
should change, too.  However, as it depends on NEI effects, etc.,
modeling its time variation is much harder and is beyond the scope of
this paper.  Without estimates of the time variation for thermal
emission, we leave it open whether the rate of the time
variation over 8 yrs is at $\sim$4\% or less than 1\%, or whether
the entire effect is due to calibration uncertainties.

\section{Discussion}

We have investigated time variations of discrete regions in the NE
limb of SN~1006, using two {\it Chandra} observations taken in 2000
and 2008.  We found that there are no particular features showing
strong time variations, and that the synchrotron emission stays at
constant within 4\% and probably with 1\% over the time span.  This
result distinguishes SN~1006 from core-collapse SNRs such as
RX~J1713.7-3946 and Cas~A in which several hot spots show year-scale 
time variations of a factor $\sim$2 or more (e.g., Uchiyama et al.\
2007).  To understand the cause of the difference between SN~1006 and
others, it should be noted that there are no knotty features in the
SN~1006 NE limb.  In fact, diffuse regions in RX~J1713.7-3946 and
Cas~A do not show fast time variations, either.  This suggests that
rapid time variations are only observed in bright knotty features.
Such a situation is indeed predicted by a recently proposed theory
that interactions between SNR shocks and ambient small-scale cloudlets 
amplify magnetic fields through plasma instabilities, and resultant
strongly magnetized features (which appears as knots or filaments)
show rapid brightness changes (Giacalone \& Jokipii 2007; Inoue et
al.\ 2009).  In this view, the fact that we do not find knotty
features which could show rapid time variations in SN~1006 is
reasonably interpreted as it is located at high Galactic latitude
where small-scale cloudlets are not present.  Additionally, SN 1006
as a Type Ia remnant is interacting with undisturbed ISM instead of
the stellar wind of the progenitor as is likely for the other two
objects, and massive-star winds may be quite clumpy.  Further
investigations for the rest of the limb of SN~1006 and other SNRs will 
be good opportunities to test the scenario for the origin of rapid
time variations in terms of amplification of magnetic fields.

Our failure to find strong time variability in the synchrotron
emission from SN~1006 is consistent with the absence of small
structures in its morphology.  While significant brightness changes on
a timescale of a few years may be explained as electron acceleration
or synchrotron-loss timescales, requiring magnetic field strengths of
0.1 -- 1 mG (e.g., Uchiyama et al.\ 2007), the absence of such changes
does not require that the magnetic fields be weak.  Steady-state
particle acceleration at the shock, followed by downstream convection
in the presence of energy losses, would result in synchrotron flux
varying only on the timescales estimated in the Appendix A, which are
independent of $B$ and depend only on the shock deceleration rate.
The absence of strong variability in SN~1006 may then be explained by
its being a remnant of a Type Ia supernova, expanding into relatively
uniform material.  The high magnetic fields estimated assuming filament
thicknesses are set by synchrotron losses ($B \sim 100 \ \mu$G, e.g.,
Vink \& Laming 2003; Morlino et al.~2010; Ksenofontov et al.~2005)
are not in contradiction with our result of little flux variability.

We also revealed spatial structures of the synchrotron emission in 
unprecedented detail, and found a correlation between the flux and the 
cut-off frequency.  Given that the flux likely depends on the magnetic
field, the simplest explanation is that the cut-off frequency depends
on the magnetic field as well, so that the magnetic field controls
spatial structures of both the flux and the cut-off frequency.  This
is important in understanding the mechanism limiting the maximum
energy, $E_\mathrm{max}$, of accelerated particles in SN~1006.  If the 
SNR age and/or escape of particles limit $E_\mathrm{max}$, then
$E_\mathrm{max} \propto B$ (Reynolds 2008).  In this case, the cut-off
frequency, which is proportional to $E_\mathrm{max}^2 B$, goes as
$B^3$.  On the other hand, if radiative losses limit $E_\mathrm{max}$,
then $E_\mathrm{max} \propto B^{-0.5}$, canceling the $B$-dependence
of the cut-off frequency.  Therefore, the possible $B$-dependence of
the cut-off frequency we found suggests that synchrotron radiative
losses do not limit $E_\mathrm{max}$ in the SN~1006 NE limb.  This
would mean that the observed $E_\mathrm{max}$ of electrons would apply
to ions as well.  Using the highest cut-off frequency of 
$\sim2\times10^{17}$Hz at the outermost regions and a magnetic field
just behind the shock of 90$\mu$G (Morlino et al.\ 2010), we estimate
$E_\mathrm{max}$ to be
$\sim$12($\nu_\mathrm{cutoff}/2\times10^{17}$Hz)$^{0.5}$($B$/90$\mu$G)$^{-0.5}$
TeV.  We note that it is also possible that some additional physical
effect, for  instance a dependence on the obliquity angle between the
shock velocity and upstream magnetic field, affects both electron  
injection and acceleration rate.  In the presence of such an effect,
radiative losses might still be the operative limitation on the
electron spectrum.  Another interpretation for the
correlation between the flux and the cut-off frequency is discussed 
in the Appendix B.

\section{Conclusion}

We tracked time variations in synchrotron flux of discrete regions in 
the SN~1006 NE limb from two {\it Chandra} observations in 2000 and
2008.  Unlike core-collapse SNRs RX~J1713.7-3946 and Cas~A where
year-scale variations were found in small-scale knotty structures
(e.g., Uchiyama et al.\ 2007; Patnaude \& Fesen 2009), we found that
the X-ray emission from the SN~1006 NE limb is quite steady.  We set
the upper limit of global time variations in the NE limb to be 4\% and
most likely 1\% over 8\,yrs.  While simple considerations lead to a
prediction of a decline of 1 -- 2\% over this period, calibration
uncertainties are also of comparable size.  We also revealed detailed
spatial structures of the synchrotron emission. We found a
correlation between the flux and the cut-off frequency, which suggests
that the maximum energy of accelerated electrons is not limited by
synchrotron losses.  If this is the case, the maximum energy for
electrons, which we calculate to be $\sim$12\,TeV, would be the same
as that for ions.  The correlation might also point to new physical
effects on electron injection or acceleration.  In conclusion, we
found no indications of particle acceleration or synchrotron losses in
discrete features in the SN~1006 NE limb.

\acknowledgments

We acknowledge helpful scientific discussions with Una Hwang.  We are
grateful to Paul Plucinsky and Alexey Vikhlinin for discussion of the
{\it Chandra} ACIS calibration.  S.K.\ is supported by a JSPS Research
Fellowship for Research Abroad, and in part by the NASA grant under
the contract NNG06EO90A.  P.F.W.\ acknowledges the support of the NSF
through grant AST 0908566.

\section{Appendix A}

We can estimate the expected rate of change of X-ray synchrotron flux
from SN 1006 with a very simple model, emission from a homogeneous
region just behind the shock whose synchrotron radiation is produced
by a power-law distribution of electrons with an exponential cutoff at
an energy $E_{\rm max}$.  We shall assume that the shock puts a
constant fraction of post-shock energy density into relativistic
electrons and another constant fraction into magnetic-field energy.
As the shock decelerates, these energies decrease, resulting in a
decrease in the synchrotron emissivity at low energies but also a drop
in $E_{\rm max}$, giving a faster rate of decrease at photon energies
produced by electrons with $E > E_{\rm max}$.  As the remnant radius
$R$ increases, however, the intensity along a line of sight $I_\nu$
grows as $R$.  Of course the true situation is much more complex,
but these simple considerations allow for an estimate.
 
The synchrotron emissivity from an exponentially truncated power-law
distribution of electrons $N(E) = KE^{-s} e^{-E/E_{\rm max}}$ between
$E_l$ and $E_h > E_{\rm max}$ is given approximately by
\begin{equation}
j_\nu = c_j(\alpha) K B^{1 + \alpha} \nu^{-\alpha} \exp(-\sqrt{\nu/\nu_c})
\end{equation}
where $\alpha = (s - 1)/2)$ and $\nu_c \equiv c_1 E_{\rm max}^2 B$
($c_1 \equiv 1.82 \times 10^{18}$ cgs; $c_j(0.6) = 3.48 \times 10^{-12}$).
In general, $c_j \equiv c_5(\alpha) (2c_1)^\alpha$ in the notation of
Pacholczyk (1970), with $c_5(0.6) = 1.17 \times 10^{-23}$.
The intensity along a line of sight is $I_\nu = \int j_\nu\, dl
\cong j_\nu L \propto j_\nu R$.  We take $\alpha = 0.6$, roughly
the radio value, although the results are not highly sensitive
to $\alpha$.

We consider a spherical evolving supernova remnant of radius $R
\propto t^m$ and shock speed $u_s \equiv dR/dt = mR/t \propto t^{m -
1}$, expanding into a uniform medium of density $\rho$.  We assume
that the shock puts a constant fraction of post-shock thermal energy
($\propto \rho u_s^2$) into relativistic electrons:

\begin{equation}
 u_e \equiv \int_{E_l}^{E_h} N(E) dE 
  \cong {K \over {s - 1}} \left( E_l^{1-s} - E_{\rm max}^{1 - s} \right)
\cong {K \over {s - 1}} E_l^{1 - s}
\end{equation}
where we have assumed $E_l \ll E_{\rm max}$.  Then if $E_l$ = const.,
$K \propto u_e \propto u_s^2 \propto t^{2m - 2}$.
Next we assume that the magnetic energy density $B^2/8\pi$ is
amplified to a (probably different) constant fraction of $\rho u_s^2$:
$B^2 \propto u_s^2 \Rightarrow B \propto u_s \propto t^{m - 1}$.

For energies far below the cutoff energy $E_{\rm max}$ (i.e., at observing
frequencies $\nu \ll \nu_c$), we can find the time-dependence of the
intensity $I_\nu$ along any given line of sight:
\begin{equation}
I_\nu \propto j_\nu R \propto t^{2m - 2} t^{(m - 1)(1 + \alpha)} t^m 
= t^{(m - 1)(3 + \alpha) + m}.
\end{equation}
For SN 1006, in the NE, $m = 0.54$ (Paper I).  
Then for $\alpha = 0.6$, 
\begin{equation}
I_\nu (\nu \ll \nu_c) \propto t^{(-0.46)(3.6) + m} = t^{-1.12} \equiv t^p.
\end{equation}

Then the prediction for the decay of synchrotron emission below
$\nu_c$, for instance in the radio, is
\begin{equation}
{1 \over I_\nu} {dI_\nu \over dt} = {p \over t} = -0.11\% \ {\rm yr}^{-1}, 
\end{equation}
or a total drop of 0.89\% in 8 years.

However, as we are considering the 1--8 keV continuum, and our fitted
values for $h\nu_c$ are typically below 1 keV ($\nu_c < 2.4 \times
10^{17}$ Hz), we need to consider the time-dependence of $\nu_c$,
i.e., of $E_{\rm max}$.  If acceleration is limited by synchrotron
losses, $E_{\rm max} \propto B^{-1/2}u_s$, and $\nu_c \propto E_{\rm
max}^2 B \propto u_s^2 \propto t^{2m - 2}.$ In any case, let $\nu_c =
\nu_0 (t /t_0)^n$.

Then, writing $I_\nu = I_0 (t/t_0)^p e^{-\sqrt{\nu/\nu_c(t)}}$,  
\begin{equation}
{dI_\nu \over dt} 
 =  I_\nu \left( {p \over t} + {n \over 2t} \sqrt{\nu \over \nu_c}\right).
\end{equation}
For SN 1006, $\nu_c \sim 2 \times 10^{17}$ Hz in the synchrotron-bright NE,
so taking a mean photon energy of about 4 times that (3.3 keV), and
using $n = 2m - 2 = -0.92$ and $p = -1.12$ as above,
\begin{equation}
{1 \over I_\nu} {d I_\nu \over dt} = {1 \over t} (p + n) = {-2.04 \over 1000}
\Rightarrow -0.20\% \ {\rm yr}^{-1}
\end{equation}
or about --1.6\% over 8 years.  A more careful integration over the
curved spectrum between 1 and 8 keV shouldn't change this estimate by
much.

If, alternatively, the acceleration is limited by the finite age of
SN~1006, we have 
$E_{\rm max} \propto B u_s2 t \Rightarrow \nu_c \propto B3 u_s4 t2
\propto t^{7(m - 1) + 2} = t^{-1.22}$ and
\begin{equation}
{1 \over I_\nu} {d I_\nu \over dt} = {-2.34 \over 1000} \Rightarrow
-0.23\% \ {\rm yr}^{-1}
\end{equation}
or about --1.9\% over 8 years. 

For completeness, a third alternative is the escape of particles above
some energy, perhaps due to absence of MHD waves to scatter them.
Then, if waves disappear above some wavelength $\lambda_m$, 
$E_{\rm max} \propto \lambda_m B \Rightarrow \nu_c \propto \lambda_m^2
B^3 \Rightarrow n = 3(m - 1)$ (ignoring possible evolutionary changes
to $\lambda_m$).  This gives $n = -1.38$ and
\begin{equation}
{1 \over I_\nu} {d I_\nu \over dt} = {-2.5 \over 1000} \Rightarrow
0.25\% \ {\rm yr}^{-1}
\end{equation}
or about --2.0\% over 8 years.  

\section{Appendix B}

Now the acceleration rate is proportional to the
diffusion coefficient $\kappa = \lambda_{\rm mfp}c/3$, where the mean
free path $\lambda_{\rm mfp}$ is normally taken to be proportional to
the gyroradius, $\lambda_{\rm mfp} = \eta r_g = \eta E/eB$ (the last
equality applying in the extreme-relativistic limit).  In the
``quasi-linear'' approximation, $\eta = (\delta B/B)^{-2}$, where
$\delta B$ is the magnitude of resonant MHD fluctuations.  (The ``Bohm
limit'' is $\eta = 1$ or $\lambda_{\rm mfp} = r_g$; constant $\eta$ at
some other value $> 1$ is termed ``Bohm-like.''  Constant $\eta$
corresponds to a ``white noise'' spectrum of MHD waves, equal energy in
all decades of wavenumber: if $I(k) dk$ is the energy in waves with
wavenumbers in $dk$, $I \propto k^{-1}$.)  While most workers assume
Bohm-like or Bohm-limit diffusion (e.g., Berezhko, Ksenofontov, \&
V\"olk 2009), one could imagine a departure from this assumption; to
produce the correlation we observe, it would be necessary to have
$\eta$ decrease with $B$ (i.e., one needs more rapid acceleration
where the field is stronger).  To our knowledge, there is at present
no theoretical prediction of such an effect, but it might exist.
However, it can be shown (Reynolds 2004) that the most straightforward
generalization, in which $\eta$ depends on $E$ because the turbulent
spectrum of MHD waves has a different slope, $I(k) \propto k^{-n}$
with $n \ne 1$, produces the wrong correlation.
In this case, the acceleration time to energy $E$, $\tau(E)$, obeys
$\tau(E) \propto E^\beta/B$ where $\beta = 2 - n$.  Then $\beta = 1$
is the Bohm limit.  The value of $\beta$ depends on the nature of the
scattering medium; for scattering by MHD waves with a Kolmogorov
spectrum, $ n = 5/3$ so $\beta = 1/3$, and normal turbulent spectra
are expected to be steep, $n > 1 \Rightarrow \beta < 1$.  
Equating $\tau(E)$ to the synchrotron loss time gives the maximum
energy of loss-limited acceleration $E_{\rm max}({\rm loss}) \propto
B^{-[1/(1 + \beta)]}$, and corresponding cut-off frequency $\nu_c
\propto B^{(\beta - 1)/(\beta + 1)}$.  (So for Bohm-like diffusion,
$\eta$ = constant or $\beta = 1$, there is no $B$-dependence. Models
in which turbulence is generated by the cosmic rays themselves
typically produce Bohm-like diffusion.)  If $\beta < 1$, higher $B$
lowers the cut-off frequency, producing the opposite correlation to
the one we observe.  However, it is still possible that some as yet
unknown effect produces turbulence with highest power at short
wavelengths, $n < 1 \Rightarrow \beta > 1$.  In this unlikely case,
brighter regions would be expected to have higher cut-off frequencies
as observed, due to variations in $B$.  That, or some other alteration
to the standard diffusion picture, could allow acceleration to be
loss-limited.  It must be noted, however, that increasing $\eta$ above
1 in some parts of the shock lowers the maximum energy to which
particles of any species can be accelerated there, and may impact the
ability of shocks to produce the highest-energy Galactic cosmic ray
ions.

\begin{deluxetable}{lcccc}
\tabletypesize{\scriptsize}
\tablecaption{Spectral-fit parameters for example spectra in regions A
and B}
\tablewidth{0pt}
\tablehead{
\colhead{Parameters}&\colhead{Region A}
  &\colhead{Region B}
}
\startdata
$N_\mathrm{H}$ (10$^{20}$\,cm$^{-2}$)&6.8 (fixed)&6.8 (fixed) \\ 
\hline
\multicolumn{3}{c}{{\tt srcut} component}\\
Constant factor (2008/2000)& 1.00$\pm$0.05 & 0.91$\pm$0.08 \\ 
Mean spectral index & 0.502$^{+0.004}_{-0.005}$ &
0.503$^{+0.007}_{-0.006}$ \\  
Cut-off frequency (10$^{16}$\,Hz) & 19$^{+2}_{-4}$ & 1.6$\pm$0.1 \\  
Flux at 1\,GHz (Jy\,arcmin$^{-2}$) & 0.236 (fixed)& 0.052 (fixed)\\  
\hline
\multicolumn{3}{c}{{\tt vpshock} component}\\
$kT_\mathrm{e}$ (keV) & 0.5 (fixed) & 0.5 (fixed) \\
log($n_\mathrm{e}t/\mathrm{cm}^{-3}$\,sec)&10 (fixed)&10 (fixed) \\ 
O  & 4.4 (fixed)& 4.4 (fixed)\\
Ne & 1.5 (fixed)& 1.5 (fixed) \\
Mg & 15 (fixed) & 15 (fixed)  \\
Si & 50 (fixed) & 50 (fixed) \\
$\int n_\mathrm{e}n_\mathrm{H}dl$$^\mathrm{a}$ (10$^{16}$\,cm$^{-5}$)
& 0.5$\pm$0.3 & 12$\pm$1 \\ 
\hline
$\chi^2$/d.o.f. & 132/169 & 141/104  \\
\enddata
\tablecomments{Errors indicate the 90\% confidence ranges.  Abundances
  not listed are fixed to the solar values. $^\mathrm{a}$VEM
  normalized by the region area; $dl$ is the plasma depth.  
}
\label{tab:param}
\end{deluxetable}

\begin{deluxetable}{lccccccccc}
\tabletypesize{\scriptsize}
\tablecaption{Flux ratios (2008/2000)}
\tablewidth{0pt}
\tablehead{
\colhead{}&\colhead{0.4--0.8\,keV}
  &\colhead{0.8--1.0\,keV} & \colhead{1.0--8.0\,keV}  
}
\startdata
Nonthermal & 1.028$\pm$0.012 & 0.962$\pm$0.011 & 0.964$\pm$0.006\\
\hline
Thermal (no shift) & 1.021$\pm$0.010 & 0.968$\pm$0.022 & 0.947$\pm$0.016\\
Thermal (1$^{\prime\prime}$ shift) & 1.026$\pm$0.010 & 0.975$\pm$0.021 & 0.956$\pm$0.014\\
Thermal (2$^{\prime\prime}$ shift) & 1.031$\pm$0.010 & 0.991$\pm$0.016
& 0.966$\pm$0.013\\
Thermal (3$^{\prime\prime}$ shift) & 1.035$\pm$0.011 & 0.988$\pm$0.022
& 0.972$\pm$0.012\\
Thermal (mean) & 1.028$\pm$0.010 & 0.982$\pm$0.020
& 0.961$\pm$0.016\\
\hline
Nonthermal / Thermal & 0.997$\pm$0.015 & 0.980$\pm$0.023
& 1.003$\pm$0.018\\
\enddata
\tablecomments{Errors indicate the 90\% confidence ranges. 
}
\label{tab:flux_ratio}
\end{deluxetable}

\end{document}